\documentclass[12pt]{article}
\pdfoutput=1

\usepackage{amsmath,amsfonts,amssymb,mathrsfs,graphicx,color,ifthen,times}

\newcommand{\ie}{{\it i.e.}}

\newcommand{\eg}{{\it e.g.}}

\newcommand{\cf}{{\it cf.}}

\newcommand{\eq}{Eq.}

\newcommand{\fig}{Fig.}
\newcommand{\efig}{Fig.}

\newcommand{\Ref}{Ref.}
\newcommand{\Refs}{Refs.}

\newcommand{\Tab}{Tab.}
\newcommand{\eTab}{Tab.}

\newcommand{\deltacp}{\delta_\mathrm{CP}}
\newcommand{\ldm}{\Delta m_{31}^2}
\newcommand{\sdm}{\Delta m_{21}^2}
\newcommand{\equ}[1]{\eq~(\ref{equ:#1})}
\newcommand{\figu}[1]{\fig~\ref{fig:#1}}

\newcommand{\efigu}[1]{\efig~\ref{fig:#1}}
\newcommand{\etabl}[1]{\eTab~\ref{tab:#1}}
\newcommand{\bi}{\begin{itemize}}
\newcommand{\ei}{\end{itemize}}

\title{Atmospheric Neutrino Oscillations \\ for Earth Tomography}

\author
{Walter Winter\\
\\
\normalsize{Deutsches Elektronen-Synchrotron (DESY), Platanenallee 6, D-15738 Zeuthen, Germany} \\
\normalsize{E-mail:  walter.winter@desy.de}
}

\date{}

\begin{document}

\baselineskip24pt

\maketitle 

\begin{quote}
{\bf
Modern proposed atmospheric neutrino oscillation experiments, such as PINGU in the Antarctic ice or  ORCA in Mediterranean sea water, aim for precision measurements of the oscillation parameters including the ordering of the neutrino masses. They can, however, go far beyond that: Since neutrino oscillations are affected by the coherent forward scattering with matter, neutrinos can provide a new view on the interior of the earth. 
We show that the proposed atmospheric oscillation experiments can  measure the lower mantle density of the earth with a precision at the level of a few percent, including the uncertainties of the oscillation parameters and correlations among different density layers. 
While the earth's core is, in principle, accessible by the angular resolution, new technology would  be required to extract degeneracy-free information.
} 
\end{quote}

\maketitle

\section{Introduction} 

Using neutrinos for Earth tomography is a dream much older than modern oscillation physics, see \Ref~\cite{Winter:2006vg} for a review: Early proposals exploit  the increase of the neutrino cross sections with energy, leading to significant neutrino absorption over the earth's diameter for energies larger than a few TeV~\cite{Nedyalkov:1981yy,Nedyalkov:2,DeRujula:1983ya,Wilson:1983an,Askar84,Borisov:1986sm,Borisov:1989kh,Kuo95,Jain:1999kp,Reynoso:2004dt,GonzalezGarcia:2007gg}.  While absorption tomography is conceptually appealing, a technically feasible and scientifically competitive approach to neutrino Earth tomography probably requires neutrino oscillations. 

The condensing evidence for neutrino oscillations by the Super-Kamiokande~\cite{Fukuda:1998mi}, SNO~\cite{Ahmad:2002jz}, and KamLAND~\cite{Araki:2004mb} experiments between about 1998 and 2004 was concluded with the measurement of a non-zero value of the last missing mixing angle~$\theta_{13}$ by Daya Bay~\cite{An:2012eh} and RENO~\cite{Ahn:2012nd} in 2012 -- and  was finally rewarded with the Nobel prize in 2015 for the discovery of neutrino oscillations to Takaaki Kajita (Super-Kamiokande) and 
Arthur B. McDonald (SNO). Modern neutrino oscillation facilities aim for precision measurements and are designed to measure the unknown parameters, such as mass ordering and CP violation. Since coherent forward scattering in Earth matter affects neutrino oscillations~\cite{Wolfenstein:1978ue,Mikheev:1985gs}, it can used as an alternative approach  for Earth tomography compared to neutrino absorption. It in principle allows for precision matter density measurements along the propagation path of these neutrinos~\cite{Winter:2005we,Gandhi:2006gu}, and the required energies are much lower. 
While neutrino absorption tomography can be compared to X-ray tomography, neutrino oscillation tomography has one interesting additional feature: since the quantum mechanical operators in different density layers do not commute, even the reconstruction from a single baseline (propagation distance) carries information how the structure along the propagation path is arranged~\cite{Ermilova:1988pw,Ohlsson:2001ck,Ohlsson:2001fy,Arguelles:2012nw}.

\begin{figure}[t]
 \begin{center}
  \includegraphics[width=0.5\textwidth]{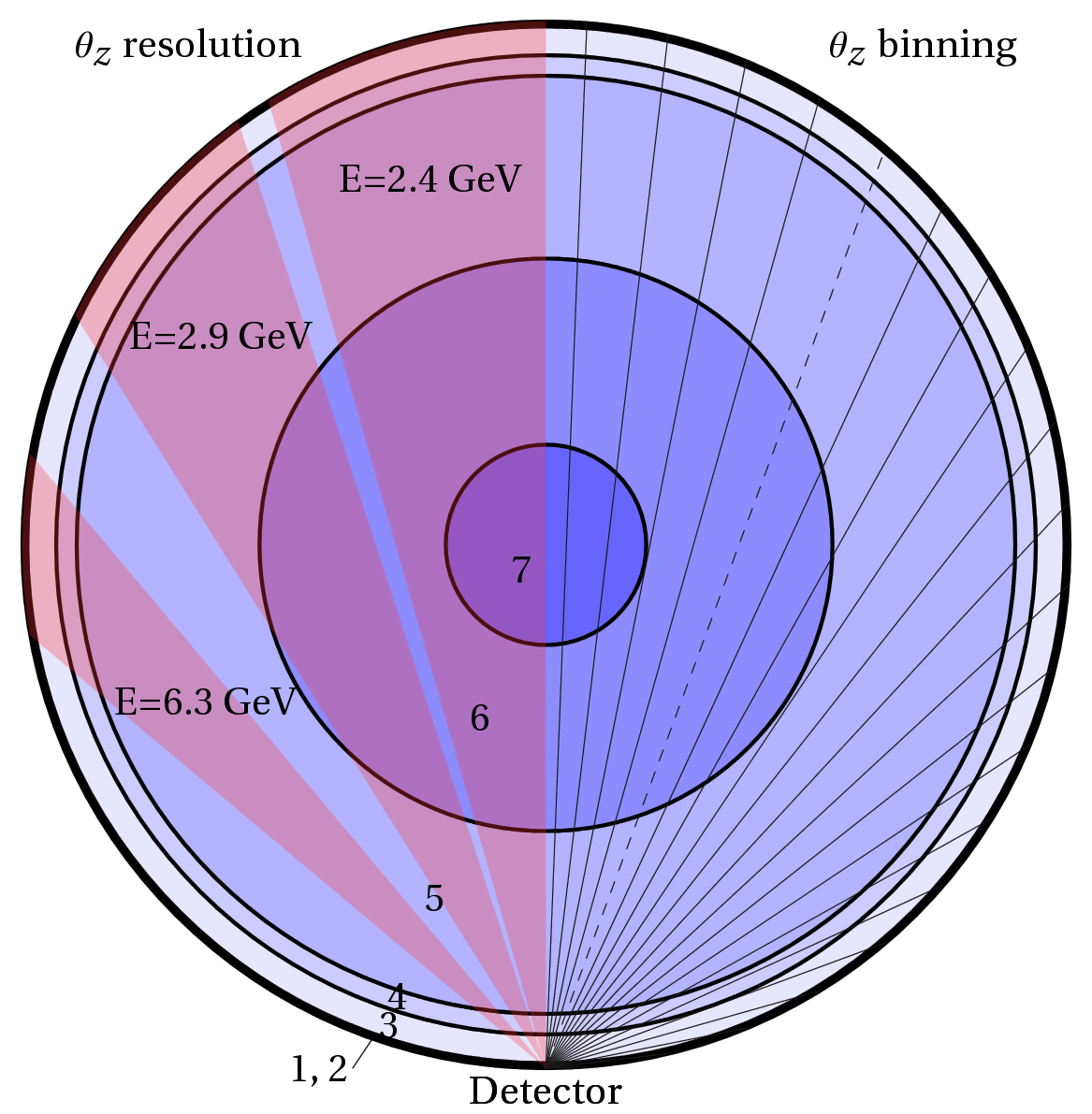}
 \end{center}
\caption{\label{fig:earthmodel} {\bf Neutrino oscillation model of the earth.} Different layers of the earth used for this analysis, adopted from the PREM model~\cite{Dziewonski:1981xy}; 1: Crust, 2: Lower Lithosphere, 3: Upper Mesosphere (mantle), 4: Transition zone, 5: Lower Mesosphere, 6: Outer core, 7: Inner core.
The right half of the figure shows the $\theta_z$ (zenith angle) binning used for the analysis, the left half of the figure illustrates the directional resolution (here for ORCA, $\bar \nu_e$~\cite{Yanez:2015uta}, $1\sigma$ range) for selected energies and directions.}
\end{figure}

Atmospheric neutrinos are produced in the earth's atmosphere by the interactions of cosmic rays continuously bombarding the earth. The generic setup, from the point of view of the detector, is illustrated in \figu{earthmodel}: neutrinos are detected from different zenith angle directions $\theta_z$ (the angle between zenith -- from the detector's viewpoint -- and incoming neutrino), which correspond to cones through the earth with different baselines $L=2 R_E \cos \theta_z$ ($R_E$: Earth radius). Within the zenith angle resolution (illustrated in left half of figure), the oscillation paths can be distinguished. We will test the structure of the earth and will identify which parts atmospheric neutrino oscillations are most sensitive within this scenario. We will use proposed experiments such as PINGU (``Precision IceCube Next Generation Upgrade'')~\cite{Aartsen:2014oha} in the Antarctic ice or or ORCA (``Oscillation Research with Cosmics in the Abyss'')~\cite{BrunnerICRC} in  Mediterranean sea water, which are modern megaton-sized  neutrino oscillation experiments designed for neutrino oscillation precision measurements with leading sensitivity to the neutrino mass ordering -- and thus the Earth matter effect; see the Appendix for the simulation techniques. 
Earlier discussions in that direction include the matter effect sensitivity~\cite{Agarwalla:2012uj} and the sensitivity to the core composition~\cite{Rott:2015kwa}. 

\section{Model and Methods}
\label{sec:earth}

We propose a whole-Earth model with seven different density layers adopted from the Preliminary Reference Earth Model (PREM) profile~\cite{Dziewonski:1981xy}, which is shown in \figu{earthmodel}, to identify the regions with highest sensitivity. 
We split the PREM profile into seven layers at depths $d$, where the characteristic density jumps occur (\cf, solid curves in \figu{rhosens}) : Crust (1), $0 \lesssim d \lesssim 35 \, \mathrm{km}$, Lower Lithosphere (2), $35 \, \mathrm{km} \lesssim d \lesssim 60 \, \mathrm{km}$, Upper Mesosphere (3),  $60 \, \mathrm{km} \lesssim d \lesssim 410 \, \mathrm{km}$, Transition zone (4),  $410 \, \mathrm{km} \lesssim d \lesssim 660 \, \mathrm{km}$, Lower Mesosphere (5), $660 \, \mathrm{km} \lesssim d \lesssim 2860 \, \mathrm{km}$,  Outer core (6), $2860 \, \mathrm{km} \lesssim d \lesssim 5151 \, \mathrm{km}$, Inner core (7), $5151 \, \mathrm{km} \lesssim d \lesssim R_E=6371 \, \mathrm{km}$ ($R_E$: Earth radius). Note that compared to seismic waves, which tend to be reflected or refracted at density jumps, neutrino oscillations are not very sensitive to structures or even strong gradients shorter than the oscillation length~\cite{Ohlsson:2001ck}, and therefore cannot resolve the density jumps precisely. Therefore it is reasonable to adopt this knowledge from geophysics.

Each baseline (see rays in \figu{earthmodel}) is separated into sections going through the density layers. Within each density layer, we follow the PREM profile~\cite{Dziewonski:1981xy}, where the matter profile is discretized into a sufficient number of steps with constant density. The oscillation probabilities are then evaluated with the evolution operator method (see \eg\ \Ref~\cite{Ohlsson:1999um}): the initial state $| \nu_\alpha \rangle$
is propagated through all matter density slices with thicknesses $x_j$ and constant densities $\rho_j$
through all crossed layers  by
\begin{equation}
\mathcal{V}(x_j,\rho_j) = e^{ - i \mathcal{H}(\rho_j) x_j} \,
\end{equation}
as the
Hamiltonian within each layer $\mathcal{H}$ is not explicitly time-dependent. The transition probability then reads
\begin{equation}
P_{\alpha \beta} = \left| \langle \nu_\beta | 
\mathcal{V}(x_n,\rho_n) \hdots \mathcal{V}(x_1,\rho_1) | \nu_\alpha \rangle \right|^2 \, .
\label{equ:pevol}
\end{equation}
Note that in general
\begin{equation}
[\mathcal{V}(x_i,\rho_i),\mathcal{V}(x_j,\rho_j ) ] \neq 0 \quad \mathrm{for} \quad  \rho_i \neq \rho_j \, ,
\label{equ:commutator}
\end{equation}
which means that the different operators do not commute and the probability will depend on the order the layers are traversed. This is an important difference to X-ray or absorption tomography, which is only sensitive to the path-integrated attenuation.

Our measured quantity in each density layer is actually a factor linearly re-scaling the density profile in this layer, as the actual density profiles for different baselines are slightly different even if they cross the same layer. Although this model is an approximation, it yields similar results for the relative matter precision compared to alternatives (such as choosing the density within each layer to be constant), but maintains accuracy of the oscillation probabilities and the oscillation measurements for the more realistic PREM profile.  For convenience, we call the measured scaling factor for layer $i$  ``$\rho_i/\bar \rho_i$'' , and depict it as error on the average matter density. 

Note that since neutrino oscillations are not sensitive to structures or changes 
shorter than the oscillation length~\cite{Ohlsson:2001ck}, additional parameters, such as multiple layers or gradients in the layers, cannot be resolved anymore beyond that level. It is clear that similar arguments apply to {\em individual} geophysical techniques, such as using the earth's free oscillation modes, see \Refs~\cite{Kenett,Masters}. As a consequence, ``structural'' information from neutrino oscillation tomography has to rely on strong density jumps (leading to interference in the probabilities) or different baselines, and ``average'' information has to rely on some knowledge from geophysics on scales shorter than the oscillation length (smoothing the density profile). Since new ways to combine neutrino oscillations with -- or compare them to -- geophysical results require further research, and atmospheric oscillation tomography is limited by the complexity from the number of parameters (oscillation parameters, systematics, and geophysics parameters), we choose the approach introduced above. 

Furthermore, note that we do not include constraints on the total mass and rotational inertia of the earth, which means that (technically speaking) some of our variations would violate these important constraints.  However, in order to include these, one needs to define a correction scheme, \ie,  which layers are corrected for density variations to maintain these constraints. One possibility has been discussed in \Ref~\cite{Winter:2005we}: Since changes of the innermost densities of the earth (\eg, inner core) influence mass and rotational inertia less that the outermost parts (where the volume is much larger), one can use small adjustments of the outer densities to compensate for large density changes in the innermost earth in spite of the higher densities there.  Since it is clear that the final result would depend on that correction scheme, and additional constraints would rather improve our result than deteriorate it (in a similar way as the free oscillation result~\cite{Kenett}), we do not consider the total mass and rotational inertia constraints in this work. An alternative (but computationally more expensive) approach would be to generate very different fit density profiles from the very beginning, and define a measure how well they fit neutrino oscillations and other potential constraints~\cite{Kenett,Ohlsson:2001ck}.

The precision on  $\rho_i/\bar \rho_i$ is obtained by minimizing the $\Delta \chi^2$ over all oscillation parameters, auxiliary systematics parameters, and the other $\rho_j/\bar \rho_j$ ($j \neq i$) simultaneously.  We also impose a 30\% external constraint on $\rho_j/\bar \rho_j$ for $j \neq i$, \ie, we assume that there is some crude knowledge on the other layer densities from geophysics and whole-Earth constraints. From the geophysics perspective, this is a very coarse constraint. From the particle physics perspective, it has the advantage that it prevents the $n$-dimensional line minimization techniques used for the analysis from falling into unphysical solutions, such as negative densities (here the penalty $\chi^2$ would exceed nine). It does not have any significant consequences for the result, except from the outer core density measurement which suffers from correlations with the inner core density.


In order to illustrate the underlying physics, consider a simple example using neutrino oscillations in constant matter density. The oscillation probability $P_{\mu e} = P(\nu_\mu \rightarrow \nu_e)$ can (neglecting contributions from solar terms)  be approximated as
\begin{equation}
P_{\mu e} \simeq  \sin^2 \theta_{23} \, \sin^2 ( 2 \tilde{\theta}_{13} ) \, \sin^2 \left( \frac{\Delta \tilde{m}_{31}^2 L}{4 E} \right)\, . \nonumber
\end{equation}
This probability is (apart from the factor $\sin^2 \theta_{23}$) just a two-flavor oscillation probability, where the fundamental parameters $\Delta m_{31}^2$ and $\theta_{13}$ are replaced by effective parameters in matter $\Delta \tilde{m}_{31}^2 =  \xi \cdot \Delta m_{31}^2$ and $\sin (2 \tilde{\theta}_{13})  =  \sin (2 \theta_{13})/\xi$
with the mapping parameter 
\begin{equation}
\xi \equiv \sqrt{\sin^2 (2 \theta_{13}) + (\cos (2 \theta_{13}) - \hat{A})^2} \nonumber
\end{equation}
 and the matter potential $\hat{A} \equiv  \pm 2 \sqrt{2} G_F n_e E/\Delta m_{31}^2$;
the different signs refer to neutrinos (plus) and antineutrinos (minus).
Here the quantity of interest is the electron density in Earth matter $n_e$, which can be converted into the matter density by $n_e = Y_e \,  \rho/m_N$ using the electron fraction $Y_e$ (number of electrons per nucleon)  and the nucleon mass $m_N$. While one has for hydrogen $Y_e = 1$,  heavier materials prefer $Y_e \simeq 0.5$ because of approximately equal numbers of protons and neutrons. We fix $Y_e = 0.5$ in this study, but one should keep in mind that one actually measures the product of $Y_e \times \rho$.\footnote{
The allowed range for $Y_e$ is actually small for typically used geophysical composition models - which implies that the composition is much harder to measure than the matter density. The reason is that heavier stable nuclei typically contain similar numbers of protons and neutrons -- as long as there is no significant hydrogen content. A well studied example in that context is the outer core, see \Ref~\cite{Rott:2015kwa}, Table~1: The values of $Y_e$ vary at the level of one percent -- which is beyond the relative precision we find in this study.} It is easy to see that the condition $\hat{A} \rightarrow \cos (2 \theta_{13})$ minimizes $\xi$, leading to effective maximal mixing.
This case is often referred to as ``matter resonance'', and can be re-cast into a condition for energy $E_{\mathrm{res}} \,  [\mathrm{GeV}] \sim 13.4 \, \cos(2 \theta_{13}) \, \Delta m^2 \, [10^{-3} \, \mathrm{eV^2}]/(\rho \, [\mathrm{g/cm^3}])$. Using typical mantle ($\rho \sim 5 \, \mathrm{g \, cm^{-3}}$), outer core ($\rho \sim 11 \, \mathrm{g \, cm^{-3}}$), and inner core ($\rho \sim 13 \, \mathrm{g \, cm^{-3}}$) densities, one obtains $E_{\mathrm{res}} \simeq 6.3 \, \mathrm{GeV}$, $E_{\mathrm{res}} \simeq 2.9 \, \mathrm{GeV}$, and $E_{\mathrm{res}} \simeq 2.4 \, \mathrm{GeV}$, respectively. These energies are perfectly covered by the atmospheric neutrino flux, and are, in principle, detectable by the discussed experiments -- although the core resonance energies are close to the threshold. The corresponding directional resolutions are illustrated in \figu{earthmodel} (left half): From this figure, one can immediately see that excellent sensitivity is expected to the Lower Mesosphere (layer~5). Although inner core and outer core can be, in principle, resolved, the corresponding data will be smeared over direction, the covered solid angle (the event rate is proportional to) is smaller, and the relevant energies are close to the experiment threshold.  While these points can be illustrated with the simple constant matter approach, the realistic matter profile of the earth leads to interesting interference effects and a parametric enhancement coming from the oscillation length matching the mantle-core-mantle structure of the earth~\cite{Petcov:1998su,Akhmedov:1998ui}, see also \Ref~\cite{Chizhov:1999az}, which are  treated numerically. Additional complications are the composition of the atmospheric neutrino flux, containing both electron and muon flavors, and the inability of the detectors to discriminate neutrinos from antineutrinos; see \eg\ \Refs~\cite{Akhmedov:2012ah,Ge:2013zua} for details. We use two event samples (muon track- and cascade-like) for the analysis, including all these effects; for analysis details, see the Appendix.

 We point out that a ``proof of principle'' for the independent extraction of the layer densities requires an experiment simulation including systematics, correlations with the oscillation parameters, and correlations among the layer densities in a self-consistent framework, see the Appendix, which is novel in this work.
The simulation techniques are based on \Ref~\cite{Winter:2013ema} using an extended version of the GLoBES (``General Long Baseline Experiment Simulator'') software~\cite{Huber:2004ka,Huber:2007ji},  which can  handle the required level of complexity.

\section{Results}
\begin{figure}[t]
 \begin{center}
  \includegraphics[width=\textwidth]{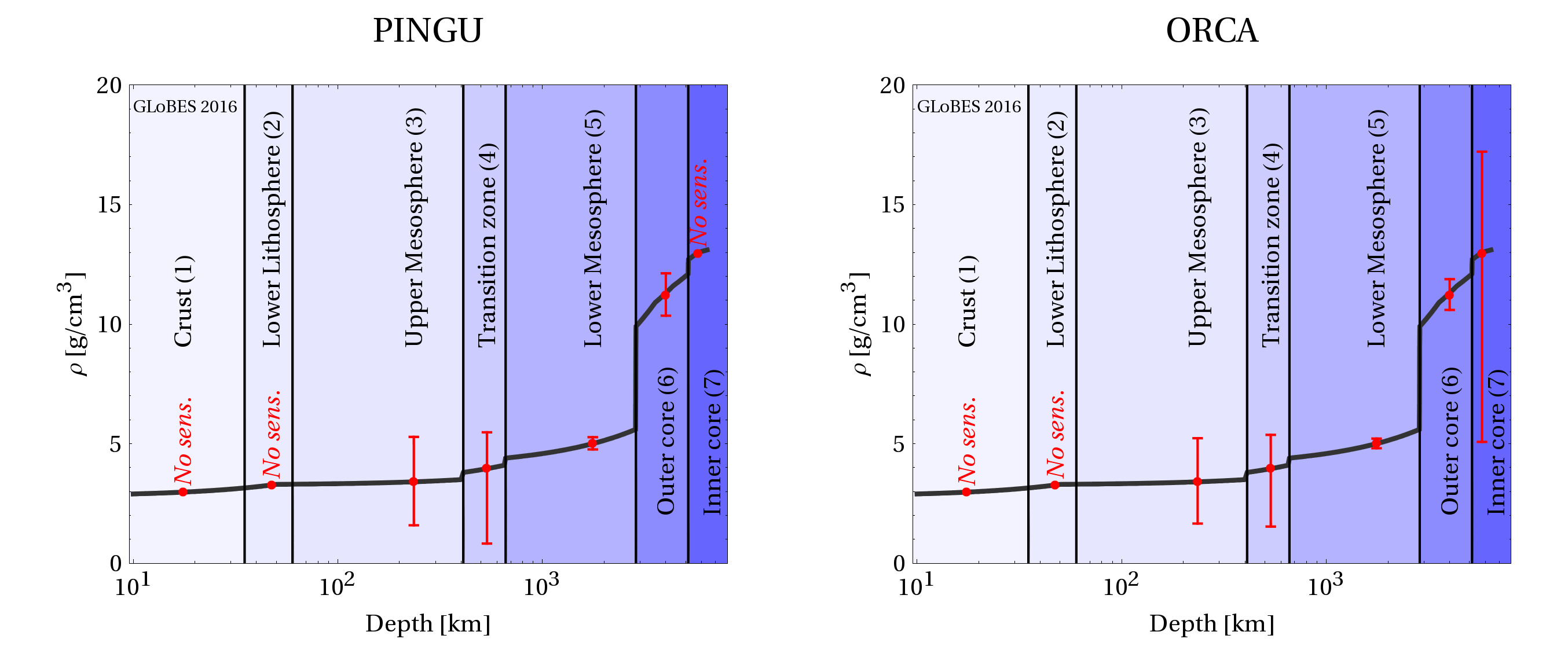}
 \end{center}
\caption{\label{fig:rhosens} {\bf Experiment sensitivity to matter density.} Projected experiment precision ($1 \sigma$ error bars) for PINGU (left) and ORCA (right) after ten years of data taking for the matter density layers corresponding to \figu{earthmodel}. Here the normal mass ordering best-fit values are assumed, and correlations (with systematics, oscillation parameters, and other layer densities) are taken into account. The solid curves correspond to the PREM matter density profile~\cite{Dziewonski:1981xy}.}
\end{figure}

\begin{table}[t]
\begin{center}
 \begin{tabular}{l|rr|rr}
\hline
& \multicolumn{2}{c|}{PINGU} & \multicolumn{2}{c}{ORCA} \\
Layer & NO & IO & NO & IO \\
\hline
Crust (1) & No sens. & No sens. & No sens. & No sens. \\
Lower Lithosphere (2) &  No sens. & No sens. & No sens. & No sens. \\
Upper Mesosphere (3) &  -53.4/+55.0  & No sens. & -51.2/+53.4 & -69.1/+52.2 \\
Transition zone (4) & -79.2/+38.3 & No sens./+72.2 & -61.2/+35.6 & -52.7/+45.8 \\
Lower Mesosphere (5) & -5.0/+5.2 & -10.5/+11.6 & -4.0/+4.0 & -4.7/+4.8 \\
Outer core (6) & -7.6/+8.2 & -40.2/No sens. & -5.4/+6.0 & -6.5/+7.1 \\
Inner core (7) & No sens. & No sens. & -60.8/+32.9 & No sens. \\
\hline  
 \end{tabular}
\end{center}
\caption{\label{tab:results} Percentage errors ($1 \sigma$) for different matter density layers
for the normal ordering (NO) and inverted ordering (IO) best-fits, including systematics and correlations with oscillation parameters and other matter layer densities.}
\end{table}

\begin{figure}[t]
 \begin{center}
  \includegraphics[width=0.9\textwidth]{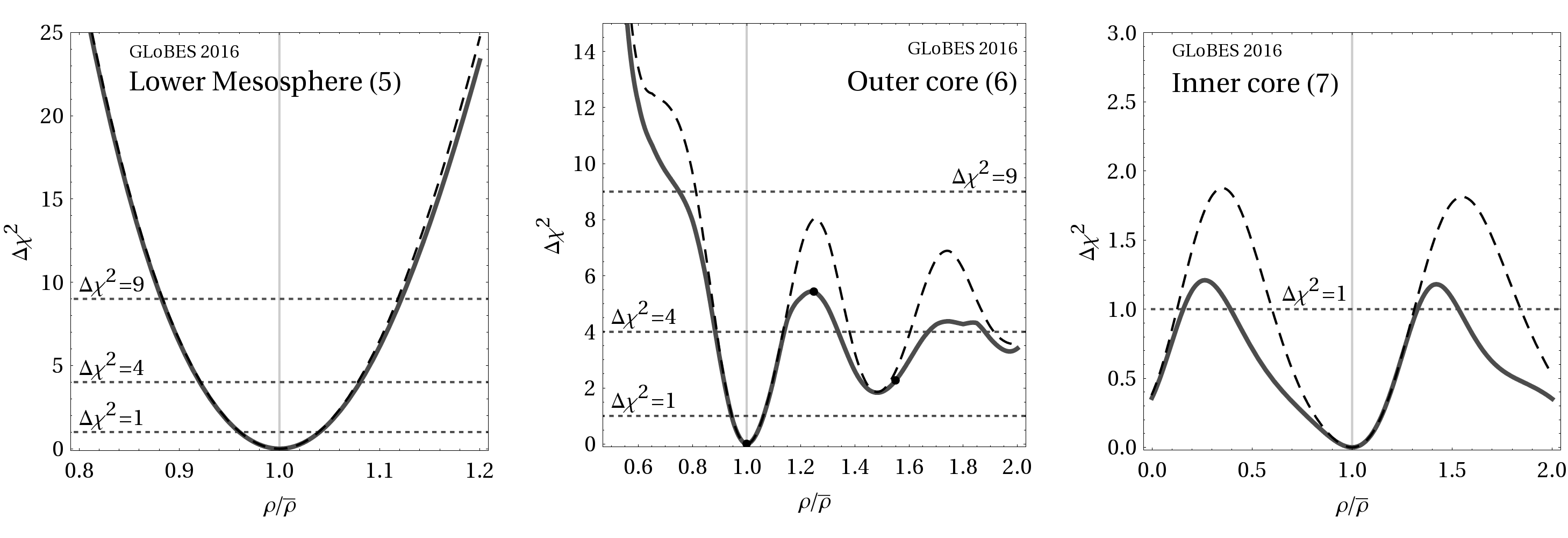}
 \end{center}
\caption{\label{fig:rholayers} {\bf Impact of parameter degeneracies.} Here the log-likelihood $\Delta \chi^2$ is shown as a function of the relative error on the average density for three different layers (in different panels) for ORCA. Solid curves include correlations among different matter density layers, with systematics, and oscillation parameters, whereas dashed curves do not include the matter density layer correlations.   The horizontal lines correspond to $1\sigma$, $2\sigma$, and $3\sigma$ for a Gaussian $\Delta \chi^2$.}
\end{figure}

For the matter density measurement, one can adopt two viewpoints: a) Tomography approach: what parts of the earth are atmospheric neutrino oscillations most sensitive to? b) Precision approach: suppose that better geophysical information exists on some layers, with what precision can a specific density be extracted? To address a), we show in \figu{rhosens} (see also \Tab~\ref{tab:results}) the expected precision including systematics and  correlations with oscillation parameters and other matter densities. Since the zenith angle resolution (\figu{earthmodel}) prohibits a resolution of layers $1$, $2$, no sensitivity can be obtained, and the sensitivity to layers $3$ and $4$ is weak. 

The best precision is found in the lower mantle with 5\% and 4\% for PINGU and ORCA, respectively. The corresponding $\Delta \chi^2$ is shown in the left panel of \figu{rholayers} for ORCA: it is well-behaved Gaussian and correlations with other density layers are not important. 
This result may, at a first glance, not be too exciting compared to the {\em collective} constraints from geophysics including free oscillations, total mass, and moment of inertia of the earth, which are believed to constrain the mantle density at the per cent level~\cite{Kenett,Masters} -- although the statistical interpretation of these precisions (confidence level of the error) seems less straightforward than in the present case. We have nevertheless demonstrated that neutrino oscillations can contribute at a similar level with an independent technique and different systematics. They may even be competitive to \Ref~\cite{Kenett} if the whole-Earth constraints (mass, rotational inertia) are included, and our method does not rely on the assumption of linearized perturbation theory as \Ref~\cite{Masters}. Future tests of neutrino tomography may use similar techniques for better comparisons, which are, however, currently subject to computational constraints.
Further applications may include the test of ambiguities and structures, such as the seismic wave-inferred low shear velocity provinces (LLSVPs) or ultra-low velocity zones (ULVZs) in the lower mantle. 

\figu{rhosens} suggests some sensitivity to the earth's outer core at $1\sigma$; however, \figu{rholayers} (middle panel) illustrates that the $\Delta \chi^2$ is not Gaussian for higher confidence levels, and degeneracies exist for $\rho/\bar \rho>1$. The difference between dashed and solid curves mainly comes from the correlation with the inner core density (the event rates mix within the zenith angle resolution).
 While the $1\sigma$ precision (solid curve) roughly corresponds to the one obtained for the core composition estimate in~\cite{Aartsen:2014oha}, it is clear that the shown degeneracies prohibit a self-consistent extraction of the outer core density up to higher confidence levels. This result  applies to the chemical composition measurement as well as, in comparison to \Ref~\cite{Rott:2015kwa},  detector setups closer to the experimental proposals are used, and the densities of the other layers are left free.

\begin{figure}[t]
 \begin{center}
  \includegraphics[width=0.5\textwidth]{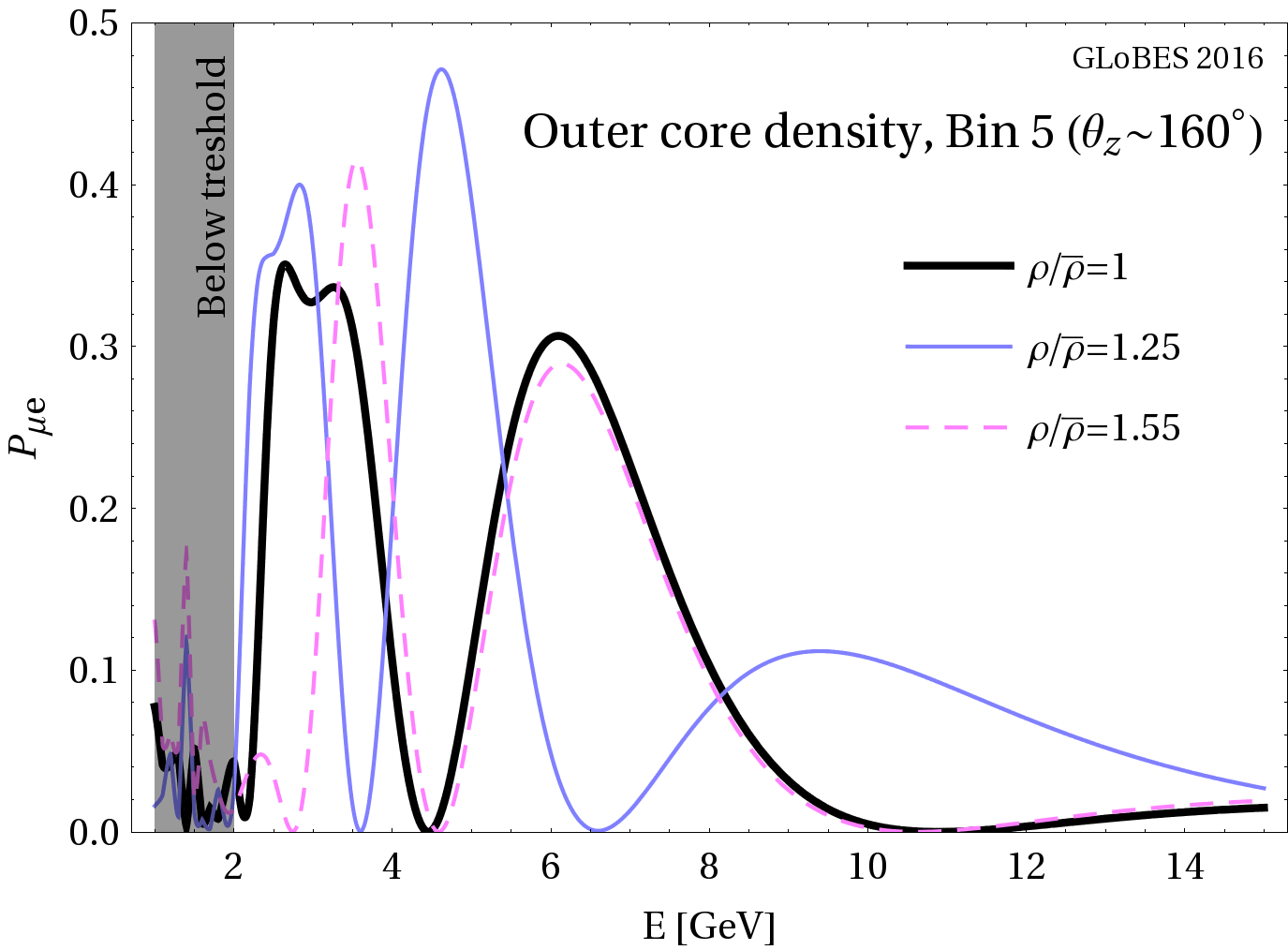}
 \end{center}
\caption{\label{fig:probs} {\bf Oscillation probability.} Oscillation probability $P_{\mu e}$ for the outer core density (6) and three different values of $\Delta \rho/\rho$ (marked by dots in the middle panel of \figu{rholayers}) and for angular bin $\theta_z \simeq 160^\circ$ (marked as dashed line in \figu{earthmodel}). }
\end{figure}

\begin{figure*}[tb]
 \begin{center}
  \includegraphics[width=0.9\textwidth]{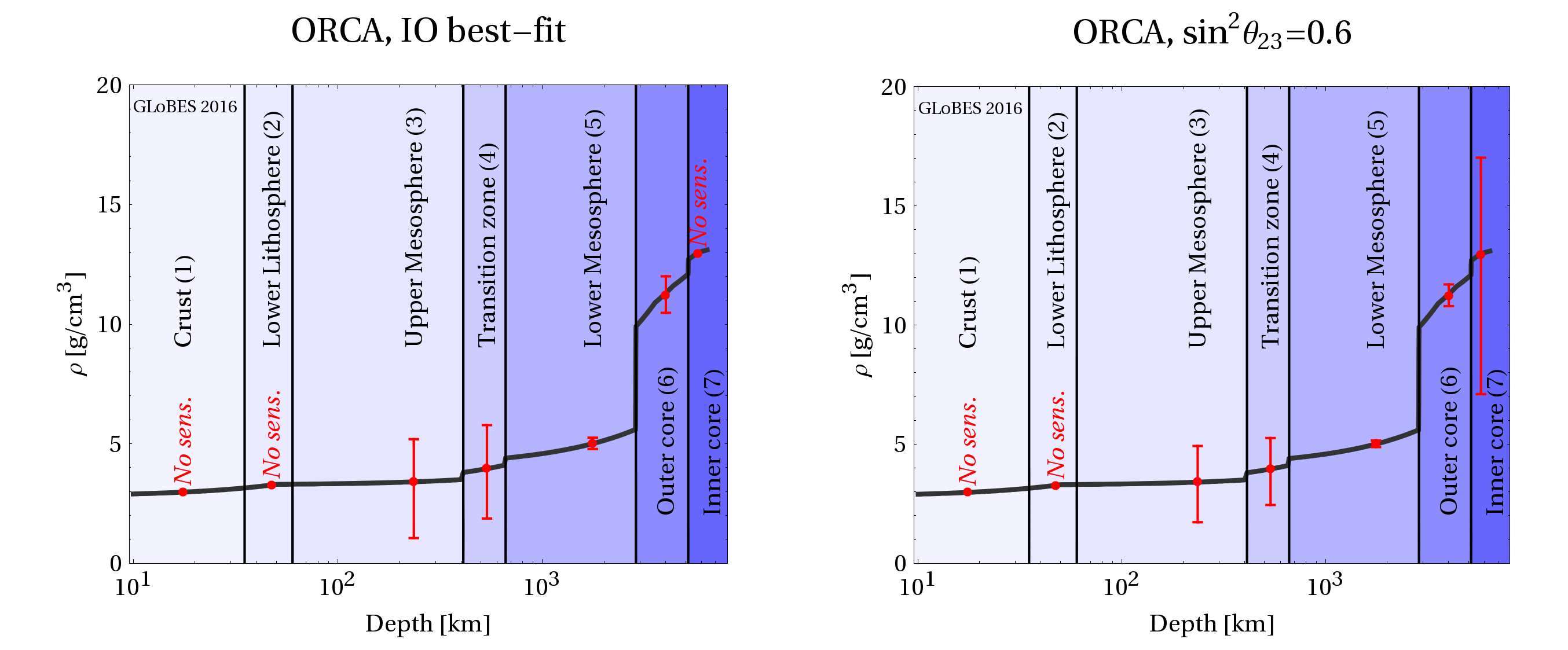}
 \end{center}
\caption{\label{fig:rhosensalt} {\bf Experiment sensitivity to matter density for alternative parameter values.} Same as \figu{rhosens} for ORCA, but for inverted mass ordering best-fit (left panel) and NO, $\sin^2 \theta_{23}=0.6$, $\deltacp=7 \pi/8$ (right panel).}
\end{figure*}

If, however, viewpoint b) is adopted, the dashed curve will represent the core density measurement, and the impact of correlations is reduced. The intrinsic oscillatory structure in \figu{rholayers} remains, as illustrated in \figu{probs} for three different values of $\rho/\bar \rho$ for $P_{\mu e}$, corresponding  to the dot marks in \figu{rholayers} (middle panel). The oscillation peak at $E \simeq 6 \, \mathrm{GeV}$ is almost perfectly re-matched by $\rho/\bar \rho = 1.55$, while it is very different for $\rho/\bar \rho = 1.25$. The low energy differences are more difficult to resolve due to the smaller effective mass and poorer directional and energy resolutions for lower energies. On the other hand, the lower bound on the outer core density is robust, 
 as lower densities correspond to higher resonance energies where the effective masses of the detectors increase. 

 We chose the NO best-fit earlier in this study; however, the actual oscillation parameters chosen by Nature may be different. We therefore show the result for ORCA and the IO in \efigu{rhosensalt} (left panel), where the performance is slightly worse (see also \Tab~\ref{tab:results}). Although the experiments include both neutrinos and antineutrinos, the antineutrino cross sections are lower -- and therefore the expected event statistics. For example, we find precision in the lower mantle of 
11\% and  5\% for PINGU and ORCA, respectively, for the IO.

Note, however,  as both the mass ordering sensitivity and the matter effect sensitivity scale with terms $\propto \sin^2 \theta_{23}$ in the appearance oscillation channels, the performance for the earth density measurements will scale in a similar way to \efigu{expall} with this parameter. This means that the actual result could be much better depending on the oscillation parameter values chosen by Nature. An example is shown in the right panel of \efigu{rhosensalt} for ORCA for a parameter set within the $3 \sigma$ currently allowed range. Here a precision of better than 3\% is obtained for the lower mantle density. In this case, the outer core density can be actually measured with a precision better than 5\%, and the degeneracies can be resolved at almost $2 \sigma$.

While the inner core density may be the prime target from the geophysics perspective, as it is the most difficult to access, currently planned instruments do not allow for a high confidence level extraction even if all the other densities were known (see dashed curve in right panel of \figu{rholayers}). This measurement operates close to the detection threshold, where also energy and zenith angle resolutions are weaker, and it suffers from a very small solid angle covered by the inner core.
A more densely instrumented detector, such as proposed in~\cite{Boser:2013oaa,Razzaque:2014vba}, would have a  lower threshold and potentially better low energy directional and angular resolutions helping both the inner and outer core density measurements. Especially in combination with geophysical data on the outer core, an extraction of the inner core density may then become possible.

\section{Summary and Conclusions}

We have demonstrated that atmospheric neutrino oscillations measured by planned detectors can provide excellent sensitivities to the lower mantle density and give a robust lower bound on the outer core density. The obtained information is complementary to that of seismic waves, as different quantities (electron density versus seismic wave velocity) and different propagation paths (straight lines versus refracted curves) are tested, and the underlying systematics are very different.  

Finally, neutrino oscillation tomography is yet a very young discipline which only has become feasible after the discovery of a non-zero value of $\theta_{13}$ in 2012. Further applications may include independent tests of irregular seismic wave propagation zones the lower mantle, where PINGU and ORCA can provide complementary information due to different locations. In the future, techniques similar to the ones used in geophysics~\cite{Kenett,Masters} may be developed, to allow for an easier comparison to and combination with geophysical data. The most inaccessible  part of the earth, the inner core, may also warrant further investigation, and could benefit from the combination with large volume detectors with lower thresholds, such as the proposed Hyper-Kamiokande~\cite{Abe:2011ts} experiment.

\subsubsection*{Acknowledgments}  I would like to thank Sebastian B{\"o}ser, J{\"u}rgen Brunner, Jannik Hofest{\"a}dt, Antoine Kouchner, Serguey Petcov, and Juan Pablo Ya{\~n}ez for useful discussions and comments, and Rolf Nahnhauer for a critical reading of the manuscript. 

 This project has received funding from the European Research Council (ERC) under the European Union’s Horizon 2020 research
and innovation programme (Grant No. 646623).


\clearpage

\begin{appendix}

\section*{Simulation Methods and Mass Ordering Sensitivity} 
\label{sec:methods}

The primary physics target for the PINGU~\cite{Aartsen:2014oha} and ORCA~\cite{BrunnerICRC} experiments is the mass ordering determination~\cite{Akhmedov:2012ah}, \ie, the question if the mass eigenstate $m_3$ is lighter or heavier than $m_1$ and $m_2$. The mass ordering can be measured by matter effects, as the resonance condition $\hat{A} \rightarrow \cos (2 \theta_{13})$ can be only implemented for neutrinos and $\mathrm{sgn}(\ldm)=+1$ (normal ordering, NO) or antineutrinos and $\mathrm{sgn}(\ldm)=-1$ (inverted ordering, IO); see definition of $\hat{A}$ in main text. Therefore, the normal mass ordering will lead to an enhancement of the oscillation effect for neutrinos and suppression for antineutrinos, and the inverted mass ordering to an enhancement for antineutrinos and suppression for neutrinos. Note that PINGU and ORCA cannot distinguish between neutrinos and antineutrinos directly, but have to rely on flux and cross section differences; the
India-based Neutrino Observatory (INO)~\cite{Ahmed:2015jtv,Indumathi:2015hfa} is a different proposal which can discriminate between neutrinos and antineutrinos by magnetization of an iron detector, but the detector mass is much smaller. Although we propose a spin-off of the main physics target in the main text, we need to establish the mass ordering determination in a self-consistent framework in order to demonstrate that the matter profile sensitivity is guaranteed even without extra equipment. 

We therefore show that we can reproduce the mass ordering sensitivities of the experimental collaborations, including the uncertainties of all oscillation parameters, such as $\deltacp$. We
treat all 6~oscillation parameters, 7~matter densities, and, using the pull method, 12~auxiliary systematics parameters equally. The precision for one parameter, such as a matter density or an oscillation parameter, can be obtained by projecting the resulting 25-dimensional fit manifold onto a one-dimensional sub-space by minimization of the $\Delta \chi^2$ over all parameters not shown.  Most importantly, this framework is fully self-consistent in the sense that any measurement of the matter density is consistent with the measurement of the oscillation parameters, which can be extracted at the same time as different projections of the fit manifold.

\subsection*{Common Simulation Framework}

\begin{table*}[t!]
\begin{center}
\begin{tabular}{p{4.5cm}rrp{5.5cm}c}
\hline
  Systematics & PINGU & ORCA  & Comments & Ref. \\
\hline
 \multicolumn{5}{l}{\bf Experiment-related systematics:} \\
Normalization & 0.25 & 0.25 & Includes  atmospheric flux normalization \\
Cross sections $\nu_\mu$, $\bar \nu_\mu$,  $\nu_e$,  $\bar \nu_e$ (CC) & 0.05 & 0.05 & Includes  uncertainty in $M_{\mathrm{eff}}$. Uncorrelated among different cross sections. & \cite{Coloma:2012ji} \\
NC normalization & 0.11 & 0.11 & Value comparable to pull obtained in recent ORCA studies & \cite{BrunnerICRC} \\[0.2cm]
 \multicolumn{5}{l}{\bf Uncertainties of atmospheric neutrino flux:} \\
Normalization & \multicolumn{3}{l}{Included in ``Normalization'' above.} & \cite{Franco:2013in} \\
Slope error (zenith bias)  &  0.04  &  0.04 & Tilt of spectrum in $\cos \theta_z$ & \cite{Esmaili:2013fva,Honda:2006qj} \\
Flavor $\nu_e$/$\nu_\mu$ & 0.01 & 0.01 & Error in flavor ratio & \cite{Honda:2006qj} \\
Polarity $\bar \nu_\mu$/$\nu_\mu$  & 0.02 & 0.02 & Error in neutrino-antineutrino ratio & \cite{Honda:2006qj} \\
Polarity $\bar \nu_e$/$\nu_e$ & 0.025 & 0.025  & Error in neutrino-antineutrino ratio  & \cite{Honda:2006qj} \\
Normalization down-going events & 0.04 & 0.04 & Value similar to zenith bias &  \cite{Honda:2006qj} \\[0.2cm]
 \multicolumn{5}{l}{{\bf Impact of Earth model:} (included if explicitly stated)} \\
Matter density & 0.3 & 0.3 & Error on matter density $\times$ composition, uncorrelated among layers~1 to~7 \\
\hline
\end{tabular}
\end{center}
\caption{\label{tab:sys} {\bf Considered independent systematical errors.}  The second and third columns list the relative errors assumed for PINGU and ORCA systematics. Altogether, there are 12~systematics pulls and 7~density parameters included in the analysis. }
\end{table*}

We simulate PINGU and ORCA for the first time within an identical oscillation framework, the same binnings,  the same systematics implementation and parameters, the same definition and computation of the performance indicators as for long-baseline experiments such as LBNF-DUNE~\cite{Adams:2013qkq}, and the same Earth density profile, using an extended version of the GLoBES software~\cite{Huber:2004ka,Huber:2007ji}. 
That GLoBES version  allows for  user-defined, channel-based systematics treatment across experiment boundaries, which was first applied in \Ref~\cite{Coloma:2012ji}. For atmospheric neutrino oscillation experiments, the different zenith angle bins are defined as different experiments in GLoBES. The directional smearing is performed after the channel-based event rate computation using pre-computed migration matrices directly compiled into the software, whereas the energy resolution is a built-in feature of GLoBES. 
The  simulation itself is an update of \Ref~\cite{Winter:2013ema}, extended by  cascade event sample and the ORCA experiment.

The energy binning for both experiments is chosen in steps of 1~GeV from $2$ to $50 \, \mathrm{GeV}$, and in steps of 10~GeV from $50$ to $100 \, \mathrm{GeV}$. The  oscillation probabilities are evaluated at a sufficiently large number of sampling points to capture fast oscillation features. For the directional binning, we choose a binning in zenith angle $\theta_z$ instead of $\cos \theta_z$, where the zenith angle bin centers correspond to the rays in \figu{earthmodel}. There is a simple reason for this choice: the $\cos \theta_z$-binning becomes coarser towards the earth's innermost parts, which we are most interested in. We have checked that the statistical difference for the mass ordering sensitivity between $\theta_z$ and $\cos \theta_z$ binning  is small. We also include two overflow bins $78.5^\circ \hdots 90^\circ$ and $0 \hdots 78.5^\circ$ corresponding to down-going events. These down-going neutrinos create a non-oscillating background for the mass ordering determination (down-going events close to the horizon may be reconstructed as up-going ones), while the overflow bins can (in principle) be used to constrain systematical errors. Since PINGU is fully contained in IceCube, we assume that the veto of atmospheric muons is good enough to use the overflow bins to constrain systematical errors, whereas we do not use the statistical information in these bins in ORCA.

The zenith angle smearing (redistribution of events) between incident $\theta_z$ and reconstructed $\theta_z'$ is computed by an oscillation-channel dependent migration matrix $R_{ijk} \equiv R(E_i, (\theta_z')^j,\theta_z^k)$ ($i$, $j$, $k$: bin indices) as a function of neutrino energy $E_i$. This implies that upgoing events exceeding $\theta_z=180^\circ$ are reconstructed in the corresponding zenith angle 
bin under a different azimuth. The advantage of this method is that the zenith angle resolution does not become asymmetric at the boundaries, and Gaussian behavior is better reproduced. We interpret the zenith angle resolutions $\Delta \theta_z$ in terms of a normalized Gaussian
\begin{equation}
 f(E, \theta_z',\theta_z) = \frac{1}{\Delta \theta_z(E)  \sqrt{2 \pi}} \exp \left( - \frac{(\theta_z-\theta_z')^2}{2 (\Delta \theta_z(E))^2} \right) \,  \label{equ:gmap}
\end{equation}
in order to compute the migration matrix
\begin{equation}
R_{ijk} \equiv R(E_i, (\theta_z')^j,\theta_z^k)  = \int\limits_{(\theta_z')_{j,\mathrm{min}}}^{(\theta_z')_{j,\mathrm{max}}} f(E_i,\theta_z',\theta_z) d\theta_z' \, 
\end{equation}
integrated over the $\theta_z'$ range covered by zenith angle bin $j$. 
The energy smearing between incident $E$ and reconstructed $E'$ energy is described by an energy smearing matrix $S_{ij} \equiv S(E'_i,E_j)$. We parameterize this matrix with a Gaussian 
\begin{equation}
g(E',E)=\frac{1}{\Delta_E(E) \,\sqrt{2\pi}}\,e^{-\frac{(E-E')^2}{2 (\Delta_E(E))^2}} \, 
\label{equ:respfun}
\end{equation} 
similar to \equ{gmap}, such that
\begin{equation}
S_{ij} \equiv S(E_i', E_j)  = \int\limits_{E'_{i,\mathrm{min}}}^{E'_{i,\mathrm{max}}} g(E',E_j) d E' \,  
\end{equation}
integrated over the $E'$ range of the $i$th energy bin, unless noted otherwise. 
Note that we interpret the median zenith angle and energy resolutions given by the experimental collaborations as $\Delta \theta_z(E)$ and $\Delta_E(E)$, which reproduces their sensitivities sufficiently well. This interpretation has limitations depending on the structure of the actual event migration matrices, such as non-Gaussianities, which can be only addressed by the experimental collaborations. However, it has the advantages that it can be identically applied to both experiments and allows for comparability and transparency of the assumptions.

For the simulation, we use two event samples (muon tracks and cascades), which are a combination of a number of channels. For instance, one ``channel'' corresponds to $\nu_e \rightarrow  \nu_e$ (CC), including a specific source flux, cross sections, efficiencies, and fiducial volume;
 altogether there are 32~such oscillation channels (from the two initial flavors $\nu_e$, $\nu_\mu$ into the three final flavors $\nu_e$, $\nu_\mu$, $\nu_\tau$ makes six, times two for neutrinos and antineutrinos plus four non-oscillating neutral current channels, makes 16, times two because separate channels with separate efficiencies for the two event samples are needed, makes 32).
Each neutrino event is counted either in the ``right'' channel as signal, or in the ``wrong'' channel as background, depending on  the (mis-)identification probabilities.

Systematics are treated exactly in the same way as in long-baseline simulations. It is important to note that not only the values for the systematical errors are important, but also how systematics are correlated among different channels and bins. For example, one may not know a certain cross section, but one does know that the same value has to be used everywhere the same interaction is measured. These kind of correlations are self-consistently implemented, see \Ref~\cite{Coloma:2012ji} for details. We list the considered systematics in \etabl{sys}, together with the assumed values; each of these corresponds to one auxiliary parameter. For example, the cross section $\times$ fiducial mass errors for different event types are assumed to be known (externally measured) to about 5\% in the considered energy range, see \eg\ discussion in \Ref~\cite{Coloma:2012ji}. 
Note that at this point we adopt the identical systematics implementation and errors for both experiments, following  physical arguments (\eg, cross section uncertainties); however,  one may think about alternative, more inclusive concepts, see \eg\ \Ref~\cite{Capozzi:2015bxa}.

For the atmospheric neutrino flux, we use updated versions~\cite{Athar:2012it}, where we use the azimuth-averaged solar-min version for the South Pole (PINGU) and Gran Sasso (close to ORCA) sites. The best-fit oscillation parameters are taken from \Ref~\cite{Gonzalez-Garcia:2014bfa}: $\sin^2 \theta_{12} = 0.304$, $\sin^2 \theta_{13} = 0.0218$, $\sdm = 7.50 \, 10^{-5} \, \mathrm{eV}^2$, and for the normal ordering  $\sin^2 \theta_{23} = 0.452$, $\ldm=+2.457 \, 10^{-3} \, \mathrm{eV}^2$, $\deltacp=306^\circ$ or for the inverted ordering $\sin^2 \theta_{23} = 0.579$, $\ldm=-2.449 \, 10^{-3} \, \mathrm{eV}^2$, $\deltacp=254^\circ$.
Note that these solutions lie in different octants, and that $\deltacp$ is slightly different. We take into account the current uncertainties on $\theta_{13}$, $\theta_{12}$, and $\sdm$ using external priors with the uncertainties from \Ref~\cite{Gonzalez-Garcia:2014bfa}, but we do not constrain $\ldm$ and $\theta_{23}$ externally. 

We compute the $\Delta \chi^2$ for a certain true mass ordering as the minimal $\Delta \chi^2$ over all oscillation parameters with the other mass ordering. This definition is exactly the same as the one used for long-baseline experiments, such as LBNF-DUNE. It has two implications: First of all, the symmetrical
  $(\ldm)_{\mathrm{eff}}$ for the $\nu_\mu$ disappearance channels~\cite{deGouvea:2005hk,Nunokawa:2005nx,deGouvea:2005mi,Ghosh:2012px,Agarwalla:2012uj,Blennow:2012gj}
\begin{equation}
  (\ldm)_{\mathrm{eff}}  = \ldm 
 - \sdm (\cos^2 \theta_{12} - \cos \deltacp \, \sin \theta_{13} \, \sin 2 \theta_{12} \, \tan \theta_{23}) \, ,   \label{equ:sym} 
\end{equation}
 clearly indicates that the mass ordering sensitivity must, in general, depend on the fit value of $\deltacp$. This has been demonstrated for PINGU, see Fig.~3 in \Ref~\cite{Winter:2013ema}. Second, it is well known that for non-maximal atmospheric mixing, the minimal $\chi^2$ may be either found in the wrong ordering-right octant (fit $\theta_{23}$ similar to true $\theta_{23}$) or in the wrong ordering-wrong octant (fit $\theta_{23} \simeq \pi/2 -$ true $\theta_{23}$) region. Note that some of the  published documents of the experimental collaborations do not yet include all of these effects, which are however essential for a fair assessment of the matter profile sensitivity.

\subsection*{Experiment-Dependent Specifications}

\begin{figure*}[t]
 \begin{center}
  \includegraphics[width=0.45\textwidth]{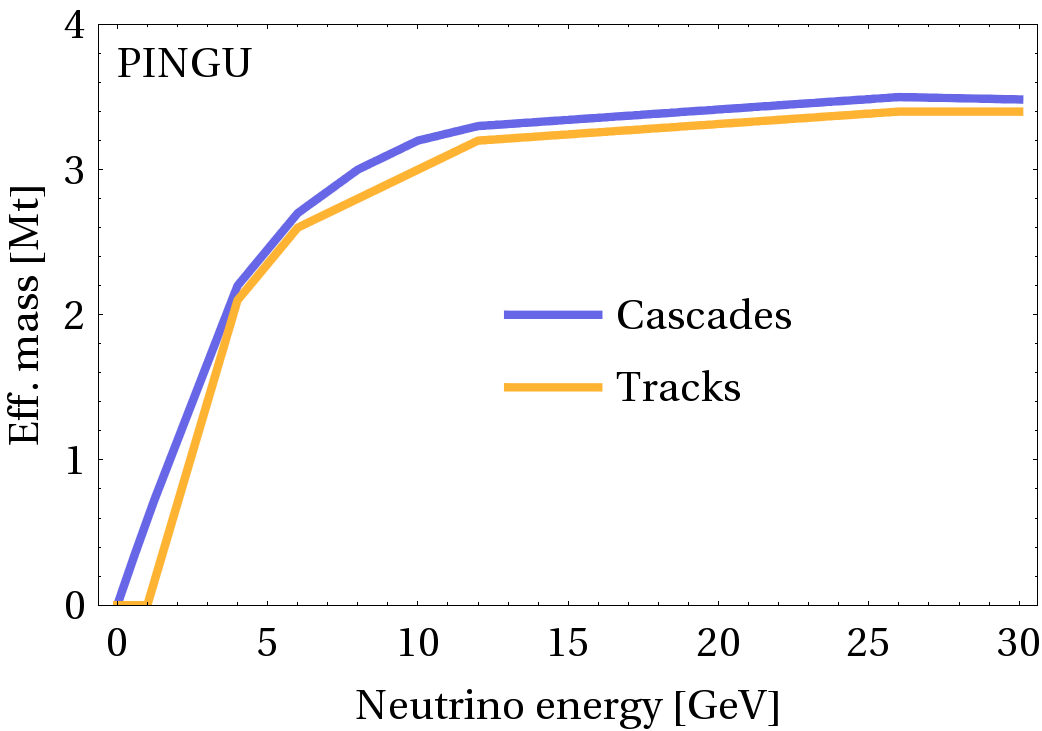} \hspace*{0.02\textwidth}%
  \includegraphics[width=0.46\textwidth]{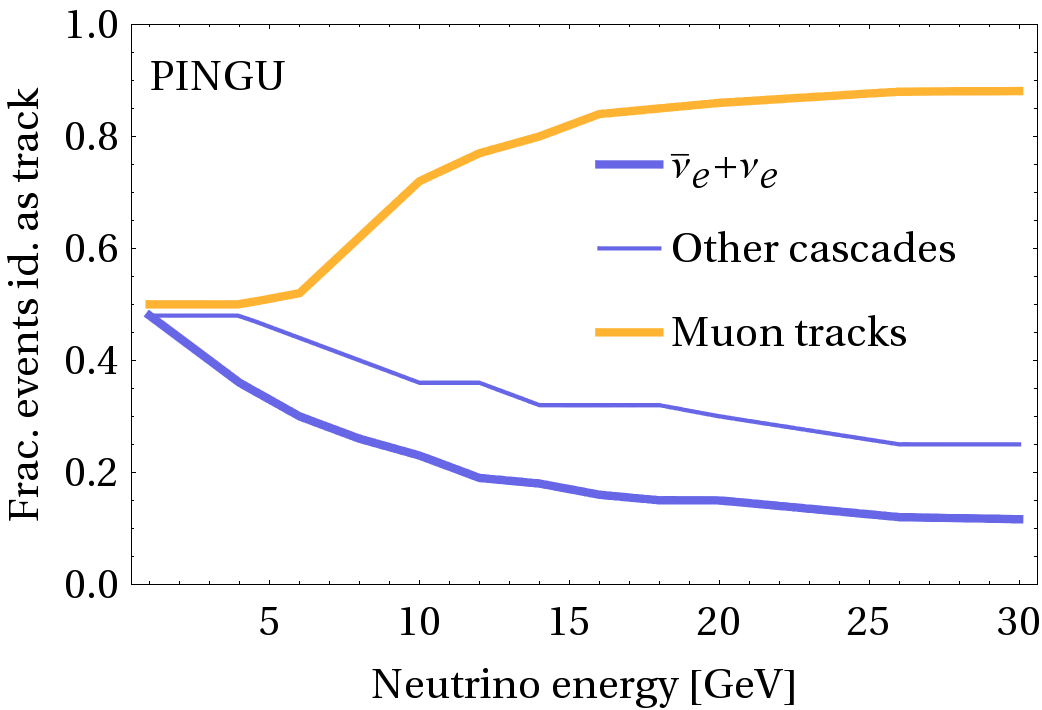} \\[0.02\textwidth]
  \includegraphics[width=0.45\textwidth]{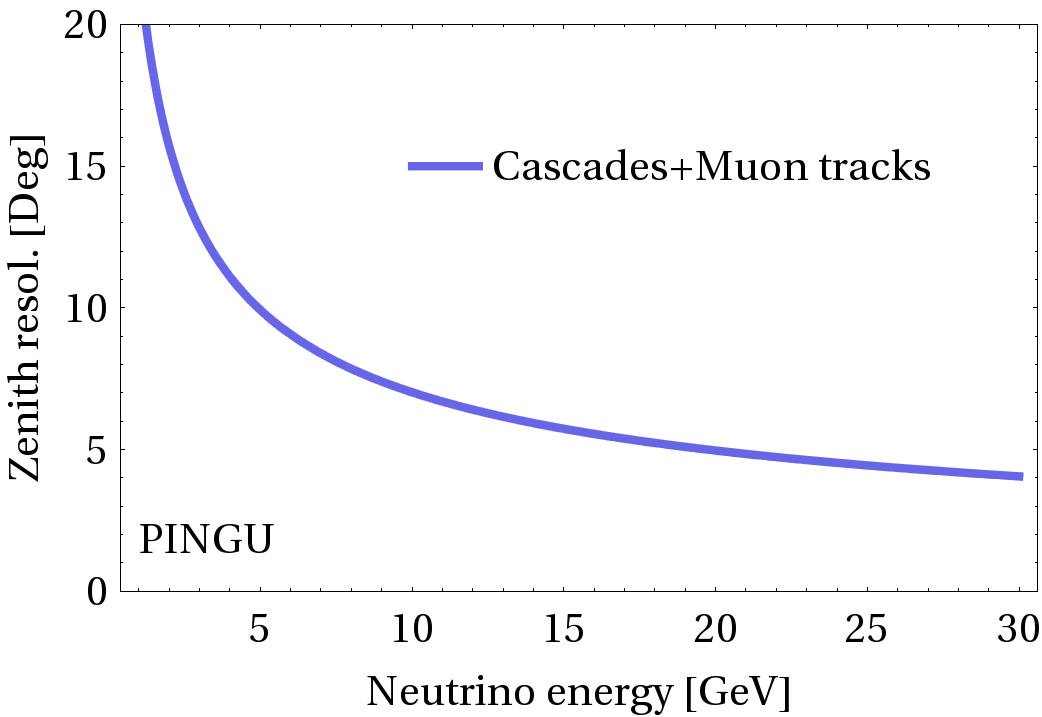}\hspace*{0.02\textwidth} %
  \includegraphics[width=0.46\textwidth]{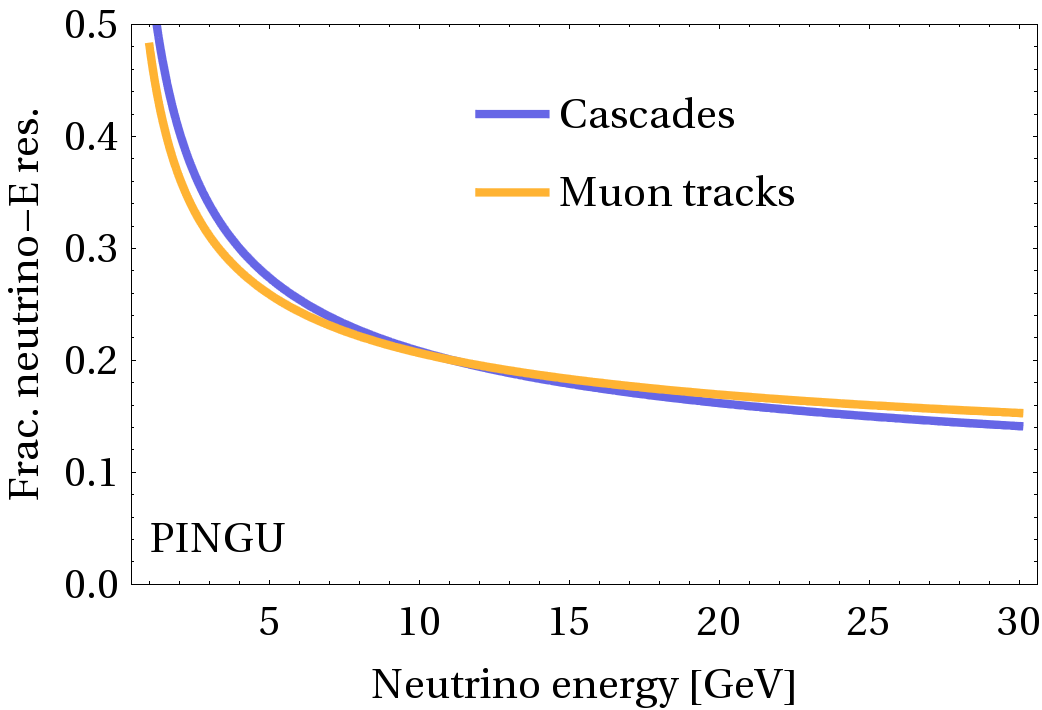}
 \end{center}
\caption{\label{fig:ppingu} {\bf Parameterization of PINGU.} Effective mass (upper left panel), fraction of events identified as track (upper right panel),  zenith angle resolution (lower left panel), and fractional neutrino energy resolution (lower right panel) for different event types (as indicated in panels) as a function of neutrino energy.}
\end{figure*}

Experiment-specific assumptions for this study include the effective mass (for different event types), the mis-identification probabilities between muon tracks and cascades, and the zenith angle and energy resolutions. 

For PINGU, we use the 40 string version documented in the Letter of Intent~\cite{Aartsen:2014oha}. The corresponding quantities are shown in \efigu{ppingu} , where we use identical parameterizations wherever the curves for different event types are very similar. The zenith angle resolution, \efigu{ppingu}, lower left panel, is parameterized by $\Delta \theta_z(E)= 0.4 \sqrt{m_p/E}$ (radians), the energy resolution, in the lower right panel, by  $\Delta_E(E)/E = 0.05 + 0.5/\sqrt{E/\mathrm{GeV}}$ and $\Delta_E(E)/E = 0.08 + 0.5/\sqrt{E/\mathrm{GeV}}$ for cascades and tracks, respectively.

\begin{figure*}[t]
 \begin{center}
  \includegraphics[width=0.45\textwidth]{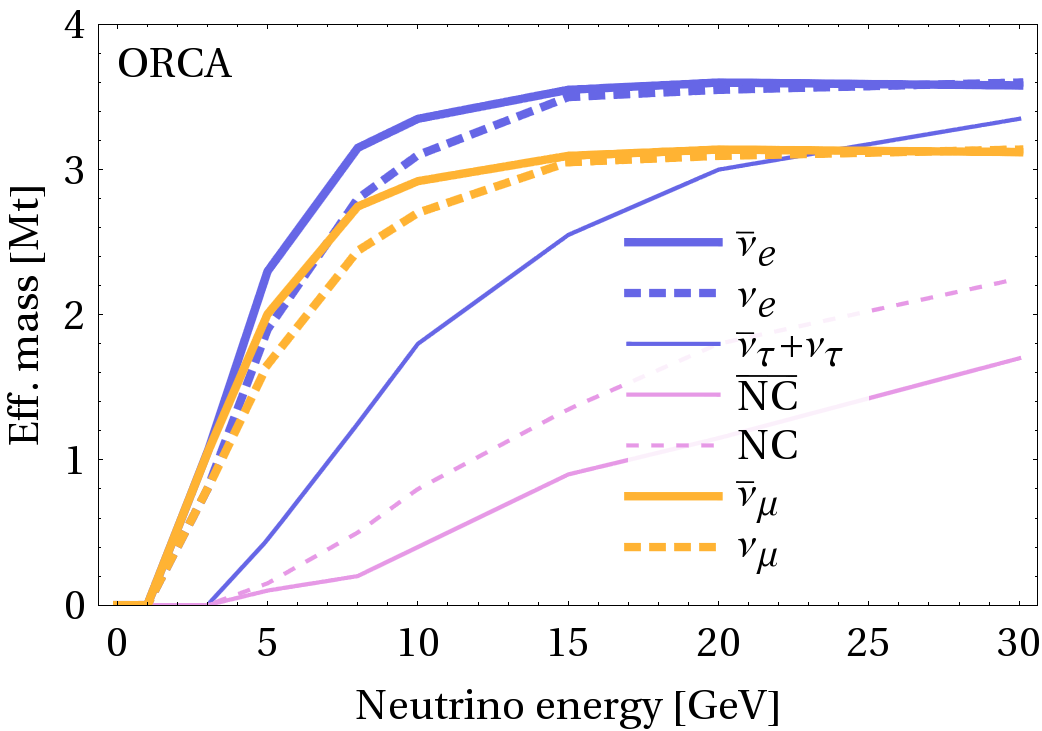}  \hspace*{0.02\textwidth}%
  \includegraphics[width=0.46\textwidth]{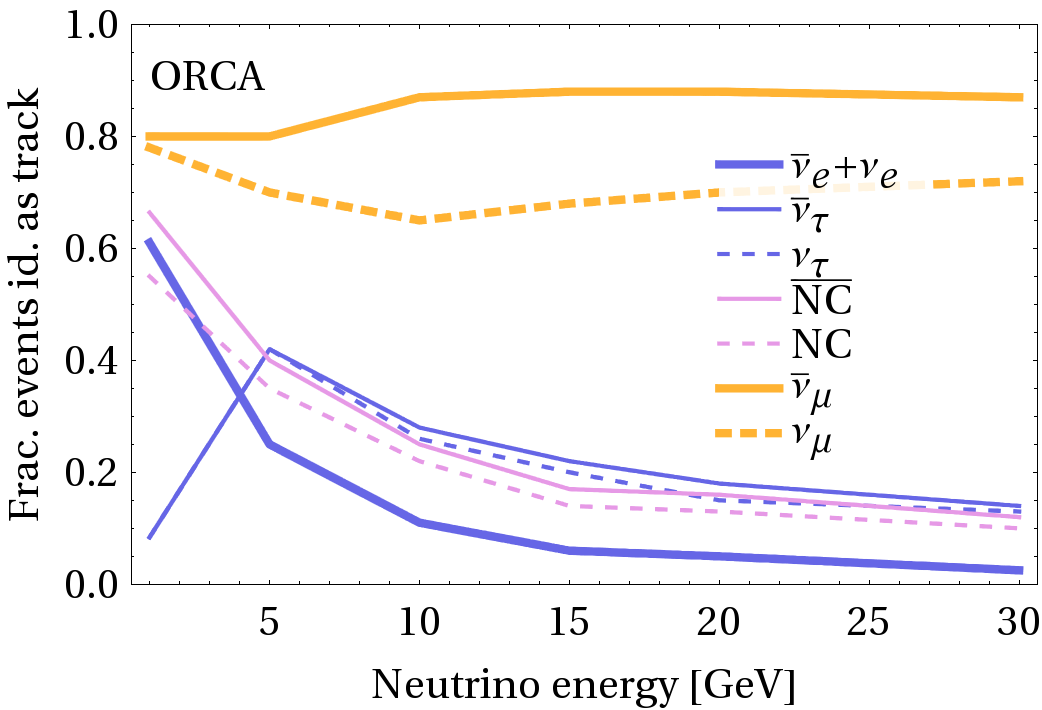} \\[0.02\textwidth]
  \includegraphics[width=0.45\textwidth]{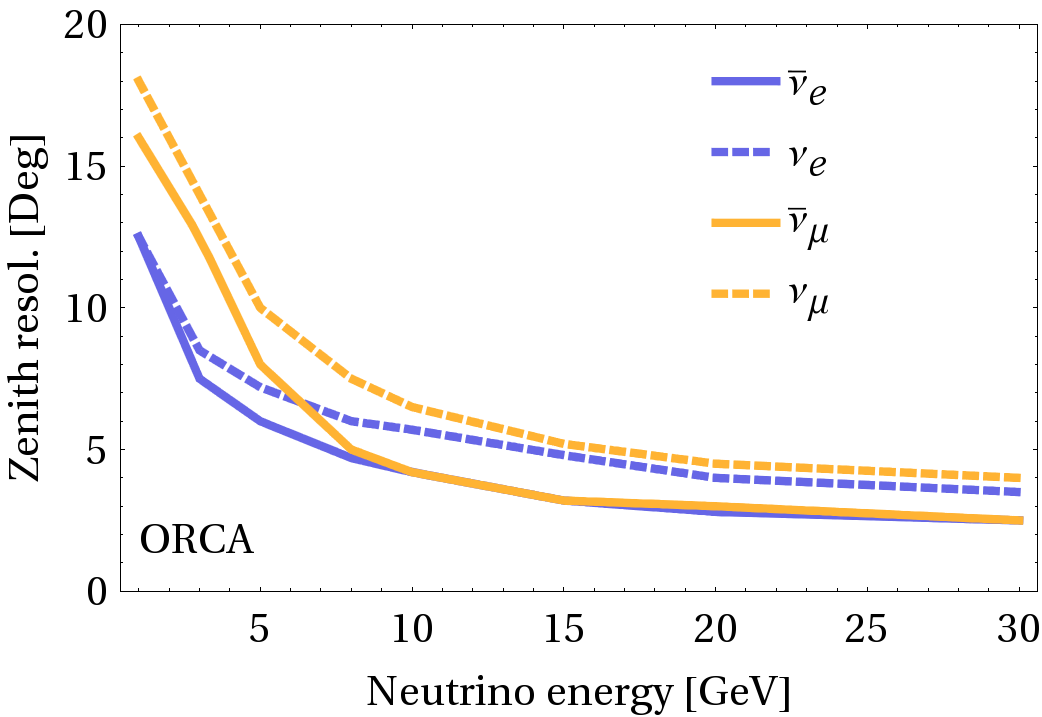} \hspace*{0.02\textwidth} %
  \includegraphics[width=0.46\textwidth]{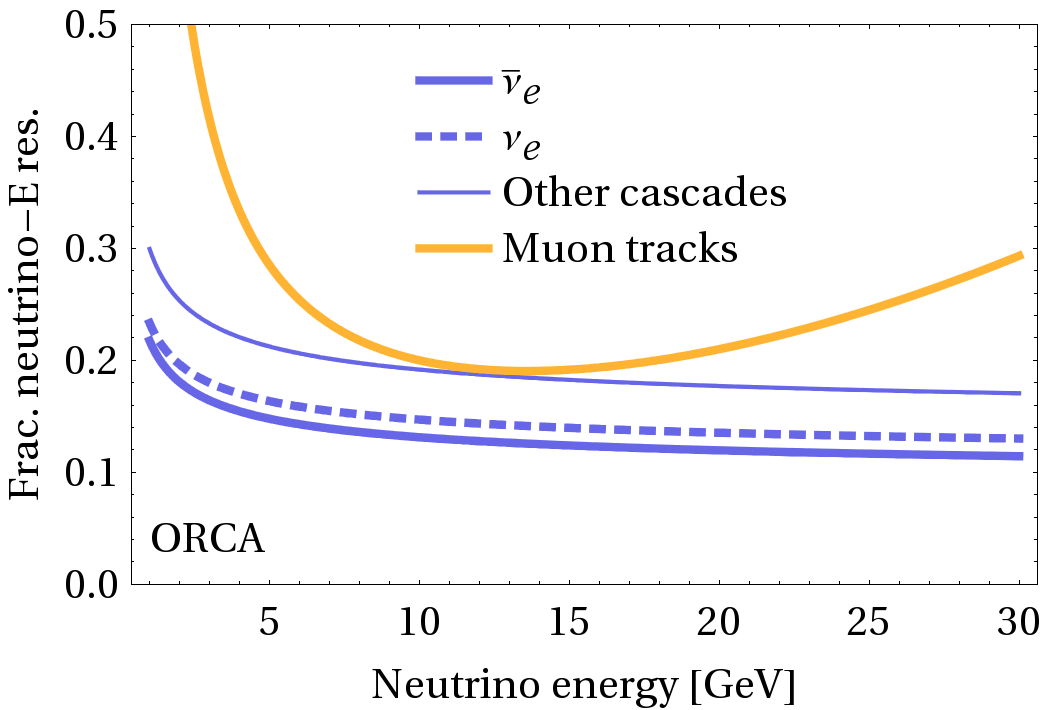}
 \end{center}
\caption{\label{fig:porca} {\bf Parameterization of ORCA.} Effective mass (upper left panel), fraction of events identified as track (upper right panel),  zenith angle resolution (lower left panel), and fractional neutrino energy resolution (lower right panel) for different event types (as indicated in panels) as a function of neutrino energy.}
\end{figure*}

For ORCA, we use the setup with 6m Digital Optical Module spacing, following the information presented at the International Cosmic Ray Conference (ICRC) 2015~\cite{BrunnerICRC,GalataICRC,JongenICRC,HofestaedtICRC}, in consistency  with \Ref~\cite{Yanez:2015uta}. The corresponding quantities are shown in \efigu{porca}. While the effective mass (upper left panel) for all cascade channels is directly obtained from \Ref~\cite{HofestaedtICRC}, the effective mass for muon tracks for this configuration has been adopted from \Refs~\cite{BrunnerICRC,GalataICRC}. The flavor identification information (upper right panel), comes from \Ref~\cite{BrunnerICRC}. The  zenith angle resolution in the lower left panel can for muon tracks be directly obtained from \Refs~\cite{BrunnerICRC,GalataICRC}, whereas the resolutions for the electromagnetic cascade channels have been adopted from \Ref~\cite{HofestaedtICRC}; these resolutions are also shown in \Ref~\cite{Yanez:2015uta}. The assumed resolutions for neutral currents and hadronic cascades are assumed to be similar to muon tracks for low energies, and about a factor of two weaker for high energies.
Note that the zenith angle resolutions are not parameterized, but instead interpolating functions are used to pre-compute the migration matrices.  The energy resolutions have been obtained in a similar way from \Refs~\cite{BrunnerICRC,GalataICRC} for muon tracks, and have been adopted from \Ref~\cite{HofestaedtICRC} for cascades, see lower right panel. We assume that they are parameterized by $\Delta_E(E)/E = 0.08 + 1/(E/\mathrm{GeV}) + 0.0002 \, (E/\mathrm{GeV})^2$, $\Delta_E(E)/E = 0.091 + 0.126/\sqrt{E/\mathrm{GeV}}$, $\Delta_E(E)/E = 0.107 + 0.126/\sqrt{E/\mathrm{GeV}}$, and $\Delta_E(E)/E = 0.142 + 0.158/\sqrt{E/\mathrm{GeV}}$
  for muon tracks, $\bar \nu_e$ cascades, $\nu_e$ cascades,  and other cascades, respectively.
A systematic offset taking into account for an offset of reconstructed and visible energy has been taken into account (for $\nu_\tau$ hadronic channels, which are dominating the $\nu_\tau$ interactions, and NC cascades). 

Comparing the lower rows between \efigu{ppingu} and \efigu{porca}, our assumptions imply slightly better zenith angle and energy resolutions for ORCA than PINGU at least for $\nu$ ($\bar \nu_e$) cascades and lower energies, which means that one may expect slightly better sensitivities for the matter density measurement -- which we find in the main text.

\subsection*{Sensitivity to Mass Ordering}

\begin{figure*}[tp]
 \begin{center}
  \includegraphics[width=0.4\textwidth]{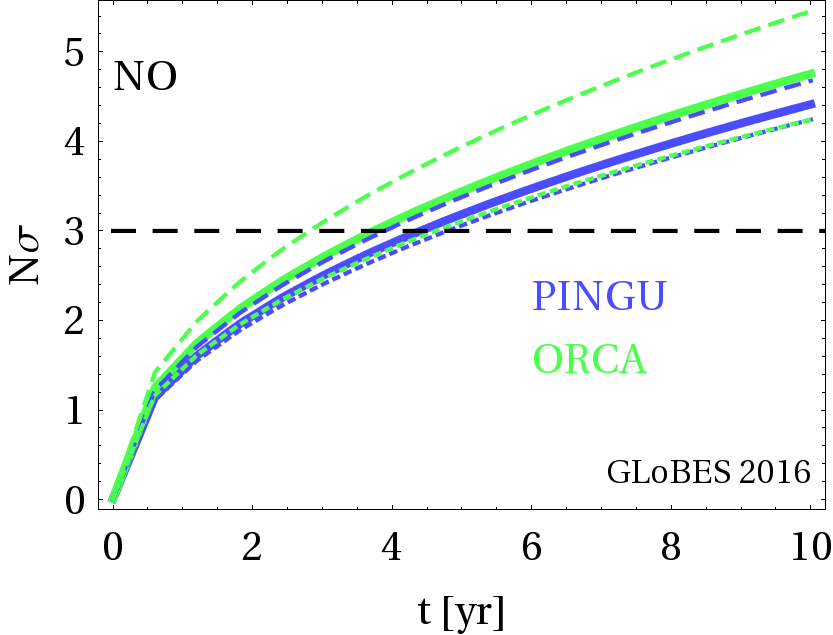} \hspace*{0.05\textwidth} %
  \includegraphics[width=0.4\textwidth]{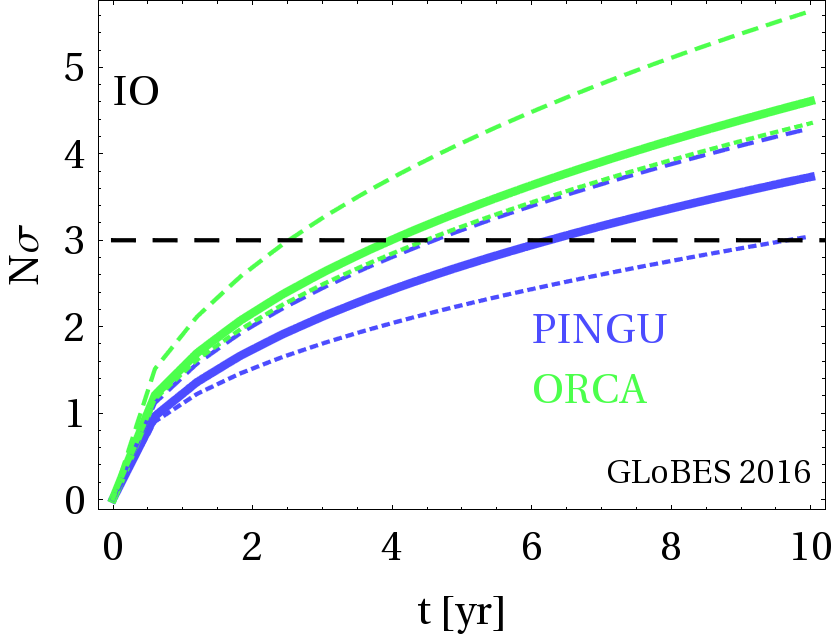}
 \end{center}
\caption{\label{fig:exposure} {\bf Mass ordering sensitivity as a function of time.} Number of $\sigma$ (assumed to be $\sqrt{\chi^2}$) for the normal (left, NO) and inverted (right, IO) mass ordering determination at the best-fit values for PINGU (blue) and ORCA (green) as a function of running time.
Solid curves include systematics and all oscillation parameter correlations and degeneracies. For a comparison to the existing literature, $\deltacp$ is fixed for the dashed curves. Dotted curves include the earth model discussed in this work as systematics, \ie, assuming unknown matter densities. }
\end{figure*}

\begin{figure*}[tp]
 \begin{center}
  \includegraphics[width=0.4\textwidth]{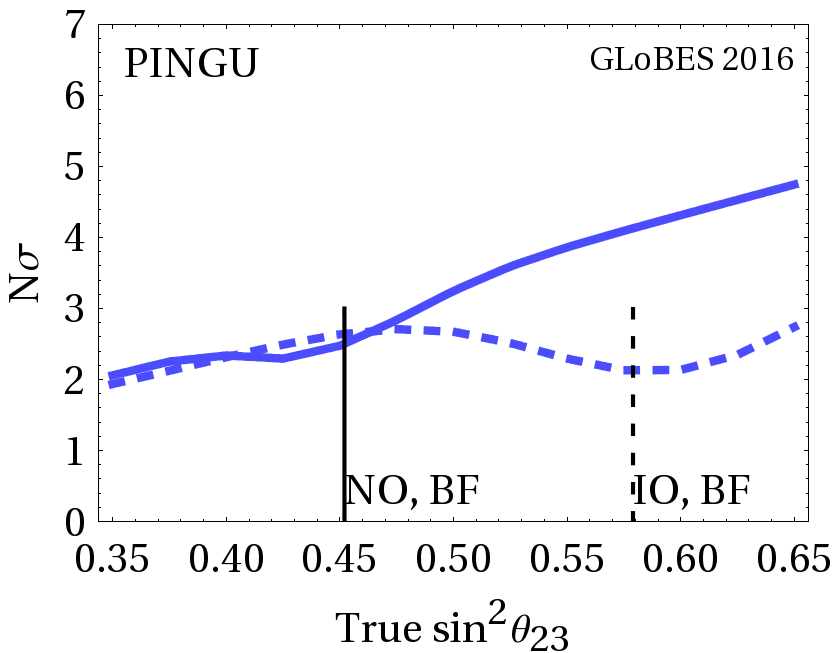} \hspace*{0.05\textwidth}  %
\includegraphics[width=0.4\textwidth]{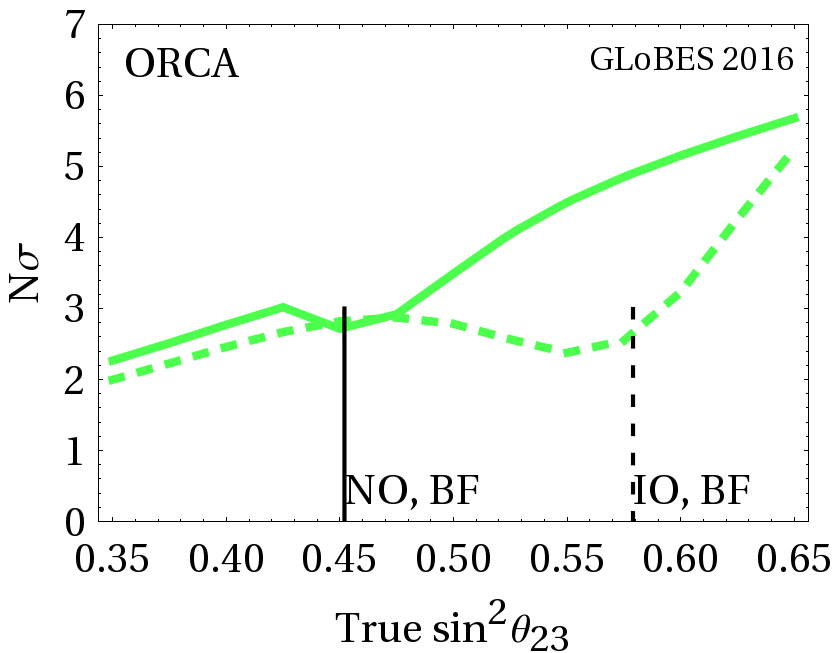}  
 \end{center}
\caption{\label{fig:t23}  {\bf Mass ordering sensitivity as a function of $\boldsymbol{\theta_{23}}$.} Number of $\sigma$ (assumed to be $\sqrt{\chi^2}$) for the PINGU (left) and  ORCA (right) as a function of the true $\sin^2 \theta_{23}$. Solid curves correspond to the normal ordering, dashed curves to the inverted ordering. The best-fit (BF) values are marked as well. Three years exposure assumed, matter densities are assumed to be known.}
\end{figure*}

\begin{figure*}[tp]
 \begin{center}
  \includegraphics[width=0.8\textwidth]{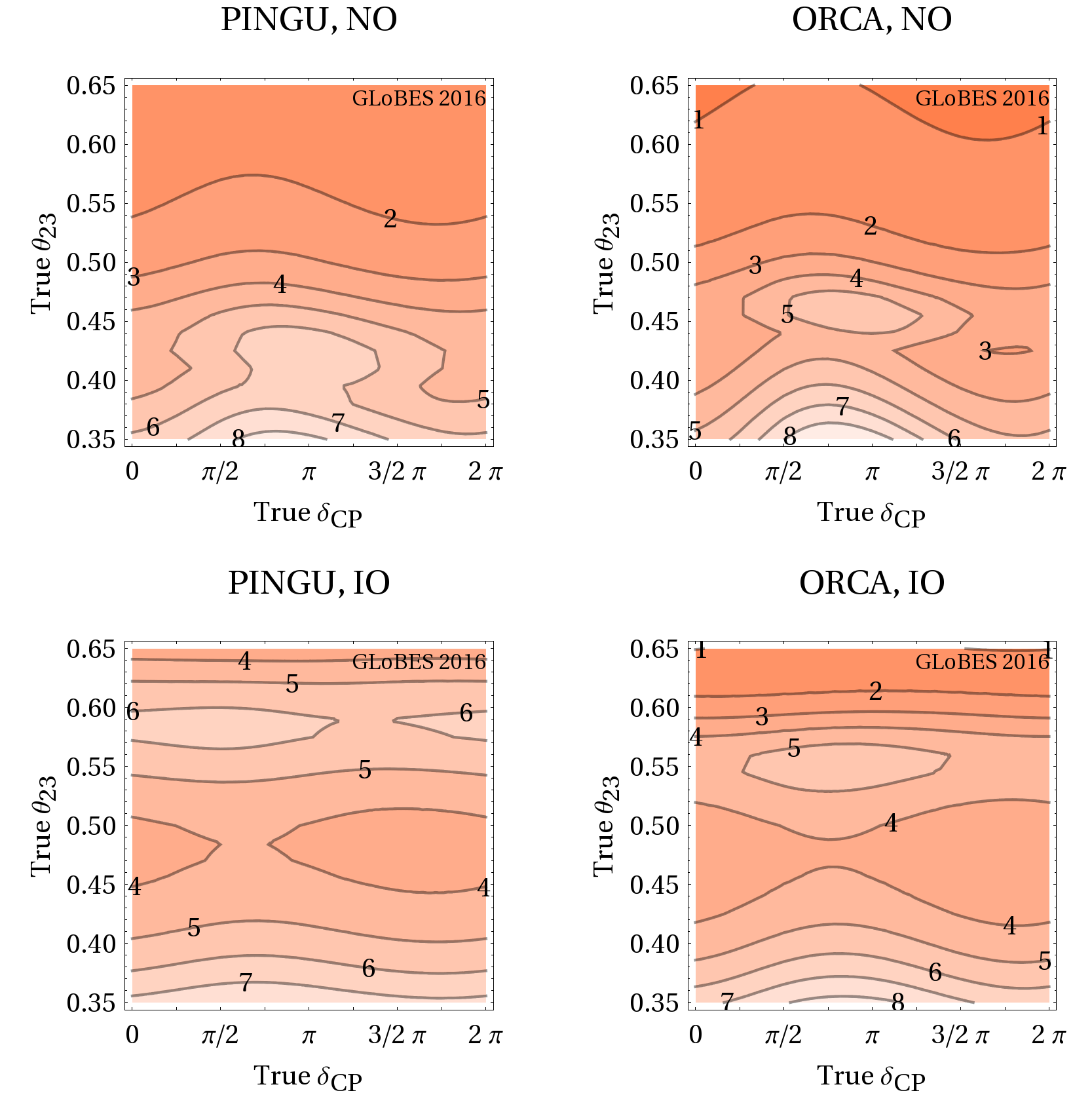} 
 \end{center}
\caption{\label{fig:expall} {\bf Required running time to establish the mass ordering at $\boldsymbol{3\sigma}$.} The contours show the approximate required running time [yr] to reach $3\sigma$ as a function of true  $\deltacp$ and true $\theta_{23}$ within the currently allowed parameter space at $3\sigma$~\cite{Gonzalez-Garcia:2014bfa}. Matter densities are assumed to be fixed.}
\end{figure*}

We show the sensitivity for the NO and IO best-fits in \efigu{exposure}. In order to compare to the existing literature, we have fixed $\deltacp$ to the respective best-fit value for the dashed curves; these reproduce the official versions from the experimental collaborations~\cite{BrunnerICRC,ClarkICRC} very well. For the solid curves, we fully take into account the minimization over $\deltacp$, which can somewhat affect the sensitivities depending on the parameter point~\cite{Winter:2013ema}.
The final sensitivities for the NO are in fact surprisingly similar for PINGU and ORCA, leading to a $3 \sigma$ discovery after about three to four years of operation within the identical systematics  and oscillation framework; for the IO the required times are somewhat longer.

While the solid and dashed curves in \efigu{exposure} assume the matter profile of the earth to be precisely known, we also perform a self-consistent simulation with the neutrino Earth model described above (dotted curves). Interestingly, we find almost identical performances for PINGU and ORCA for the normal mass ordering. The PINGU sensitivity for the inverted ordering is most affected by the unknown matter density. An inspection of the systematics pulls reveals an 18\% increase of the lower mantle density and a 6\% decrease of the outer core density (compared to the input densities) for the three year sensitivity. Here the external knowledge from geophysics may in fact be essential  to improve the mass ordering sensitivity.

An important cross-check is the dependence of the mass ordering sensitivities on the true $\theta_{23}$, see \efigu{t23}. These figures reproduce the qualitative behavior of the experimental collaborations~\cite{Yanez:2015uta} but are somewhat more conservative for ORCA because the correlation with $\deltacp$ is fully included. In some cases jumps between the octants where the minimal $\chi^2$  found, see \eg\ right panel around $\sin^2 \theta_{23} = 0.43$, which are somewhat sensitive to the experiment implementation for small $\theta_{23}$.

Finally we show in \efigu{expall} the required running time (contours, in years) to establish the mass ordering at $3\sigma$ as a function of true  $\deltacp$ and true $\theta_{23}$. For this figure,  the performance has been computed as a function of $\deltacp$ and  $\theta_{23}$ simultaneously for three years of operation. Then we  have linearly scaled the $\chi^2$ to obtain an estimate for how long it takes to reach $3 \sigma$.
 As a result, a $3\sigma$ discovery is guaranteed for PINGU and ORCA within the anticipated timescale of the matter density measurement even in remote (allowed) regions of the parameter space, whereas in the most optimistic case, ORCA can find the NO already after one year.
As a consequence, the mass ordering will be resolved in either case at the timescale of the matter density measurement discussed in the main text, and the corresponding degeneracy cannot affect these results.

\end{appendix}

\end{document}